# Guidelines for cyber risk management in shipboard operational technology systems

**Priyanga Rajaram, Mark Goh, Jianying Zhou**

iTrust, Centre for Research in Cyber Security
Singapore University of Technology and Design (SUTD)

Email: priyanga@sutd.edu.sg, mark_goh@sutd.edu.sg, jianying_zhou@sutd.edu.sg

**Abstract.** Over the past few years, we have seen several cyber incidents being reported, where some of the primary causes were the lack of proper security controls onboard the ship and crew awareness on cybersecurity. In response to the growing cyber threat landscape in the maritime sector, we have developed a set of guidelines for maritime cyber risk management, focusing on four major shipboard Operational Technology (OT) systems that are crucial for the day-to-day operation of ships. These four OT systems are: Communication Systems, Propulsion, Machinery and Power Control Systems, Navigation Systems and Cargo Management Systems. The guidelines identify the cyber risks in each of the OT systems and recommend the necessary actions that can be taken to manage risks in each shipboard OT system. In this paper, we introduce the new guidelines, which include cyber risks, mitigation measures, cyber risk assessment, and a checklist to help shipowners and maritime authorities assess and enhance cyber hygiene of their vessels. Our guidelines have been disseminated by the Maritime and Port Authority of Singapore (MPA) to owners and operators of the Singapore Registry of Ships for their reference and use.

## 1. Introduction

Cybersecurity has become a vital subject for the maritime industry to consider, due to the increasing adoption of digital advancements that help enhance maritime operations. Despite their many benefits, digitalisation of shipboard operations can pose a significant threat to the safety of vessels and its crew, if there is a lack of awareness on cybersecurity among the crew and cyber hygiene habits are not put into practice. All the components in a vessel, such as OT systems, IT systems including routers, switches and network cables, etc., need to be safeguarded from cyber threats, since a weak cybersecurity posture can have a major impact on the safety of the crew, vessel and cargo. As shipboard OT systems are crucial for day-to-day operation of vessels, managing the cyber risks associated with those systems must take top priority.

    We have developed a new set of guidelines [1] to address the impacts of cyber risks in shipboard OT systems and the necessary security controls to be enforced in those systems in order to maintain cyber hygiene in vessels. The guidelines are framed in such a way that they are feasible and cost-effective for enforcement by ship owners and maritime authorities. The guidelines are centered on four shipboard OT systems: Communication Systems, Propulsion, Machinery and Power Control Systems, Navigation Systems, and Cargo Management Systems. The International Maritime Organisation (IMO) has also highlighted these OT systems as the most vulnerable to cyberattacks [2].

    In our new guidelines, the attack surfaces and cyber risks associated with the shipboard OT systems are listed, with a description of potential attack scenarios. We also described mitigation measures for

those cyber risks identified. To assess the severity and likelihood of those cyber risks, a cyber risk assessment methodology is proposed. Finally, a checklist with a tiered security approach is included to guide maritime authorities in determining the cyber hygiene of vessels. It can also be referenced by those in charge of ensuring and enhancing cybersecurity of vessels, such as engineers and IT/OT system specialists. During each stage of developing the guidelines, webinars were hosted to exhibit the work done and to receive feedback from the stakeholders, which helped in fine-tuning the guidelines further.

The rest of this paper is organised as follows: Section 2 reviews the related guidelines that have been published by various maritime organisations. Section 3 investigates the cyber risks and attack surfaces associated with the shipboard OT systems. Section 4 highlights the features of our guidelines, especially the cyber risk assessment approach and the checklist with tiered security. Finally, Section 5 concludes the paper.

**2. Related Guidelines**

The International Maritime Organization (IMO) resolution MSC.428(98) *Maritime Cyber Risk Management in Safety Management Systems* states that a ship's Safety Management System (SMS) must include cyber risk management based on the objectives and functional requirements of the International Safety Management (ISM) code [3]. Before the first annual verification of company's Document of Compliance on 1 January 2021, IMO encourages administrators to address the cyber risks in SMS. Correspondingly, this resolution highlights IMO's *Guidelines on Maritime Cyber Risk Management* (MSC-FAL.1/Circ.3), which contains high-level recommendations on maritime cyber risk management, to safeguard the maritime industry from the existing and emerging cyber threats [2]. This guideline recommends five functional elements of cyber risk management defined by the National Institute of Standards and Technology (NIST) cybersecurity framework, that is, identify, protect, detect, respond, and recover [4]. The guidelines also emphasize on the fact that cyber risk management should be initiated by the senior management level, and awareness on cybersecurity must be established at all levels of an organisation.

Besides references to ISO/IEC 27001 and NIST cybersecurity frameworks, the IMO guideline also refers the *Guidelines on Cyber Security Onboard Ships Version 4 issued in 2020,* published and supported by various maritime organisations such as the Baltic and International Maritime Council (BIMCO), Intercargo, Intermanager, INTERTANKO, Chamber of Shipping of America (CSA), Digital Containership Association (DCA), ICS, IUMI, OCIMF, Sybass and World Shipping Council [5]. This guideline focuses on creating awareness on cybersecurity onboard ships and guides in enhancing the safety and security of the ships and the crew. It also outlines the cyber risks and the right measures to be implemented to ensure cybersecurity onboard vessels.

Det Norske Veritas (DNV) recommended a set of cybersecurity measures with the publication of *Cyber Security Resilience Management for Ships and Mobile Offshore Units in Operation in 2016,* to guide the personnel in charge of cybersecurity in their respective organisations [6]. This guideline is designed based on well recognised guidelines issued by organisations such as IMO and BIMCO. The guidelines spotlight various cyber threat factors targeting the maritime sector and focuses on crucial improvements and validation mechanisms necessary for cybersecurity resilience in ships. The 'cyber secure' guideline for class notation (October 2020 edition) provided by DNV is centred on the cybersecurity of a ship's primary functions and the operational needs of ship owners [7]. The notations, Cyber Secure, Cyber Secure (Essential) and Cyber Secure (Advanced) entail ten primary functions such as navigation, propulsion, steering, power generation, etc., and are suitable where cyber risk management system is set up and enforced into prevailing procedures and systems. Apart from this, the "+" notation can be used to add additional functions for a notation.

The European Network and Information Security Agency (ENISA) published two reports regarding maritime cybersecurity. In 2011, ENISA published a report on the *Analysis of Cyber Security Aspects in the Maritime Sector,* to highlight the cybersecurity challenges in the maritime industry and provide suggestions for the same [8]. Another related report published in 2019 is on *Good Practices for Cybersecurity in the Maritime sector,* focusing on port cybersecurity, primarily aiming at those responsible for OT and IT security in the port ecosystem [9]. That report includes good practices for ensuring port cybersecurity, related to both IT systems and OT systems.

Similarly, the Institution of Engineering and Technology (IET) published a *Code of Practice Cyber Security for Ships* in 2017, to recommend good practices on developing cybersecurity assessment plans, devising mitigation measures, and managing security breaches and incidents [10]. The guideline considers the importance of implementing cyber security measures, with respect to the technological aspects in maritime systems. On the other hand, the American Bureau of Shipping (ABS) published the *Guide for Cybersecurity Implementation for the Marine and Offshore Industries, ABS Cybersafety Volume 2,* in February 2021, providing cyber security recommendations and requirements for cyber-enabled systems [11]. ABS awards vessels with four notations CS-System, CS-Ready, CS-1 and CS-2, based on compliance with the requirements given in this guide.

Based on the review of the guidelines discussed above, while they do discuss maritime cyber risk management, their intended audiences are those at the management levels. On the contrary, our guidelines focus on managing risks in shipboard OT systems, with the target audience being those responsible for implementing cybersecurity controls in their respective organisations, and they include IT/OT system specialists and engineers. Our guidelines include a list of actionable mitigation measures organised as a checklist that can serve as a guide for maritime authorities and ship owners to check and ensure the cyber hygiene of the vessels.

**3. Cyber Risks in Shipboard OT Systems**

The shipboard OT systems that are fundamental for everyday vessel operations include communication systems, propulsion, machinery and power control systems, navigation systems and cargo management systems. These systems are vulnerable to cyber attacks due to various weaknesses in the systems or in the network in which they are connected. While enhancing operational efficiency with new technologies is important, cyber risks loom large in the background. For example, shipboard OT systems need to be connected to the Internet for operations, updates, communication, etc., and such interconnectivity between OT and IT network could be an entry point for attackers to compromise the ship's systems. The attack surfaces and corresponding cyber risks associated with each OT sub-system are provided in Tables 1, 2, 3, 4. For more details, please refer to our published journal article on cyber risks in shipboard operational technology (OT) systems in [12].

*3.1. Communication Systems*

Reliable communication is crucial during events such as alerting, sending and receiving maritime safety information. Cyberattacks targeting essential communication systems such as satellite communication system (SATCOM) and VoIP (Voice Over Internet Protocol) phones can disrupt communication among ships and between ship and shore.

*3.2. Propulsion, Machinery and Power Control Systems*

The ship's propulsion, power generation and management are vital for its safe passage. The performance parameters of machineries related to propulsion and power control are mostly monitored and controlled through a computer console. At times, these systems may use the Internet for various operational purposes, and remote updates, thus they can fall victim to cyberattacks. Compromising these systems could result in issues such as engine failure, fuel tank overflow, alarm failure and disrupted power supply in the vessel.

*3.3. Navigation Systems*

Navigation systems play a major role in determining a ship's speed, position and heading, which enable safe navigation. If cyberattacks are targeted on such systems, it may result in lack of navigation data, which in turn may result in severe consequences such as ship sinking, collision, etc.

*3.4. Cargo Management Systems*

Cargo management systems are a crucial part of vessel operations including assisting in the tracking and management of cargo, and in maintaining vessel's stability in varying environmental conditions with the use of ballast water system. With vast reliability on the Internet for managing cargo operations and due to human threat factors, these onboard systems can fall victim to cyberattacks. Attackers might take advantage of the weaknesses in these systems to compromise them, which could disrupt cargo operations and trigger vessel instability.

**Table 1.** Cyber risks in Communication Systems.

| OT sub-system | Attack surface | Cyber risk |
|---|---|---|
| Satellite Communication System (SATCOM) Integrated Communication System (ICS) | Phishing emails, VSAT system/modem | Malware attack via Phishing emails, exploiting vulnerabilities in outdated VSAT software, unauthorised administrative access of vessel network, cross-site scripting, eavesdropping |
| Voice Over Internet Protocol (VoIP) | VOIP protocol (Session Initiation Protocol) | Denial of Service (DoS) attack, eavesdropping, vishing |
| Wireless Local Area Network (WLAN) | Access point (router) | Denial of Service (DoS) attack, access point tampering, eavesdropping/session hijacking |

**Table 2.** Cyber risks in Propulsion, Machinery and Power Control Systems.

| OT sub-system | Attack surface | Cyber risk |
|---|---|---|
| Engine Governor System Fuel Oil System Alarm Monitoring & Control System | USB ports in the machinery monitoring systems, serial communication network | Malware attack via USB ports, Man-in-the-middle (MITM) attack |
| Power Management System Emergency Generators and Batteries | USB ports in the power management system console, serial communication network | Malware attack via USB ports, Man-in-the-middle (MITM) attack |

**Table 3.** Cyber risks in Navigation Systems.

| OT sub-system | Attack surface | Cyber risk |
|---|---|---|
| Electronic Chart Display and Information System (ECDIS) | USB ports, serial communication network, outdated software | Malware attack via USB ports, spoofing, Denial-of-Service (DoS) attack |
| Radio Detection and Ranging (RADAR) | Ethernet switch, SMB (Server Message Block) service | Malware intrusion, Man-in-the-middle (MITM) attack |
| Automatic Identification System (AIS) | AIS software, AIS messages (VHF radio communication) | Spoofing, Replay attack, Frequency hopping attack |
| Global Positioning System (GPS) | GPS/GNSS receiver | GPS spoofing, GPS jamming |
| Dynamic Positioning System (DPS) | DP system software, GNSS Receiver | Denial-of-Service (DoS) attack, Backdoor attack, Spoofing |

| OT sub-system | Attack surface | Cyber risk |
|---|---|---|
| Global Maritime Distress and Safety System (GMDSS) | Very High Frequency (VHF) radio communication, poor/outdated software | Eavesdropping, spoofing, Denial-of-Service (DoS) attack |
| Voyage Data Recorder (VDR) | USB ports, VDR system | Man-in-the-middle (MITM) attack, Remote code execution |
| Integrated Navigation System (INS) | SMB (Server Message Block) service, Remote desktop protocol | Man-in-the-middle (MITM) attack, Remote code execution |

**Table 4.** Cyber risks in Cargo Management systems.

| OT sub-system | Attack surface | Cyber risk |
|---|---|---|
| Cargo Control Room (CCR) | Phishing emails, USB ports | Ransomware, Malware attack |
| Ballast Water System (BWS) | Phishing emails, USB ports | Phishing emails, Malware attack |

## 4. Features of New Guidelines

While the existing guidelines mainly focus on maritime cyber risk management at the management level, our guidelines are framed in such a way that they can be readily adopted by those responsible for the cybersecurity of shipboard OT systems such as engineers, IT/OT system specialists and vessel inspectors. Here we highlight the features of our new guidelines.

*4.1. Mitigation Measures*

With heightening demands for cargo transportation by sea, if vital shipboard systems are subject to cyber attacks, the company might need to face severe repercussions such as economic and reputational loss. Ship operators, ship owners, engineers, IT/OT specialists, etc., must be aware of the potential cyber risks and the necessary mitigating actions to be taken. Our guidelines include a list of specific actionable mitigation measures to each of the cyber risks in shipboard OT systems (discussed in Section 3) that can be promptly implemented on those systems.

For example, security controls can be helpful when shipboard OT systems operate in the serial network. However, they often connect to the Internet for remote updates and operational purposes. Thus, technical and procedural security measures such as firewall, Virtual Private Network (VPN), antivirus software, email security mechanisms, regular SATCOM software updates, password security mechanisms, logging, access control lists, multi-factor authentication, network monitoring, and crew awareness programs are necessary for safe operation of the Internet-facing systems, as they are one of the common entry points for various cyberattacks. On the other hand, physical security measures are needed in cases where machineries need to be handled with care, for example, engine system, power generators, etc., as mishandling can cause harm to the crew and the ship. Likewise, it is also very important to secure USB ports in shipboard systems by blocking unused ports, to avoid insider attacks. The mitigation measures in our guidelines include a mixture of technical, procedural and physical security measures that can help safeguard shipboard OT systems from potential cyber attacks.

*4.2. Cyber Risk Assessment Approach*

In order to assess the cyber risks, a risk assessment approach is also detailed in our guidelines. Risk assessment helps in determining the impact and likelihood of risks, and this in turn can help to identify the places where security controls need to be improvised or freshly implemented. As the shipboard system conditions evolve due to upgrades, installation of new software, etc., it is good to perform risk assessment frequently.

In our guidelines, a four-step scale definition is defined to assess the severity and likelihood of a cyber risk. A 4x4 risk score matrix, as shown below in Figure 1, can be used to calculate the risk score for each cyber risk in an OT system, where the risk score is the product of severity score and likelihood score. The likelihood score of a risk can be decided based on the attack complexity and attacker's ease of access to necessary resources and attack surface. On the other hand, impact or severity score of a risk can be determined by the extent of environmental harm, financial loss and loss of confidentiality, integrity and availability that can result from a cyber attack. Tables 5 and 6 depict the definitions for evaluating the likelihood and severity scores respectively.

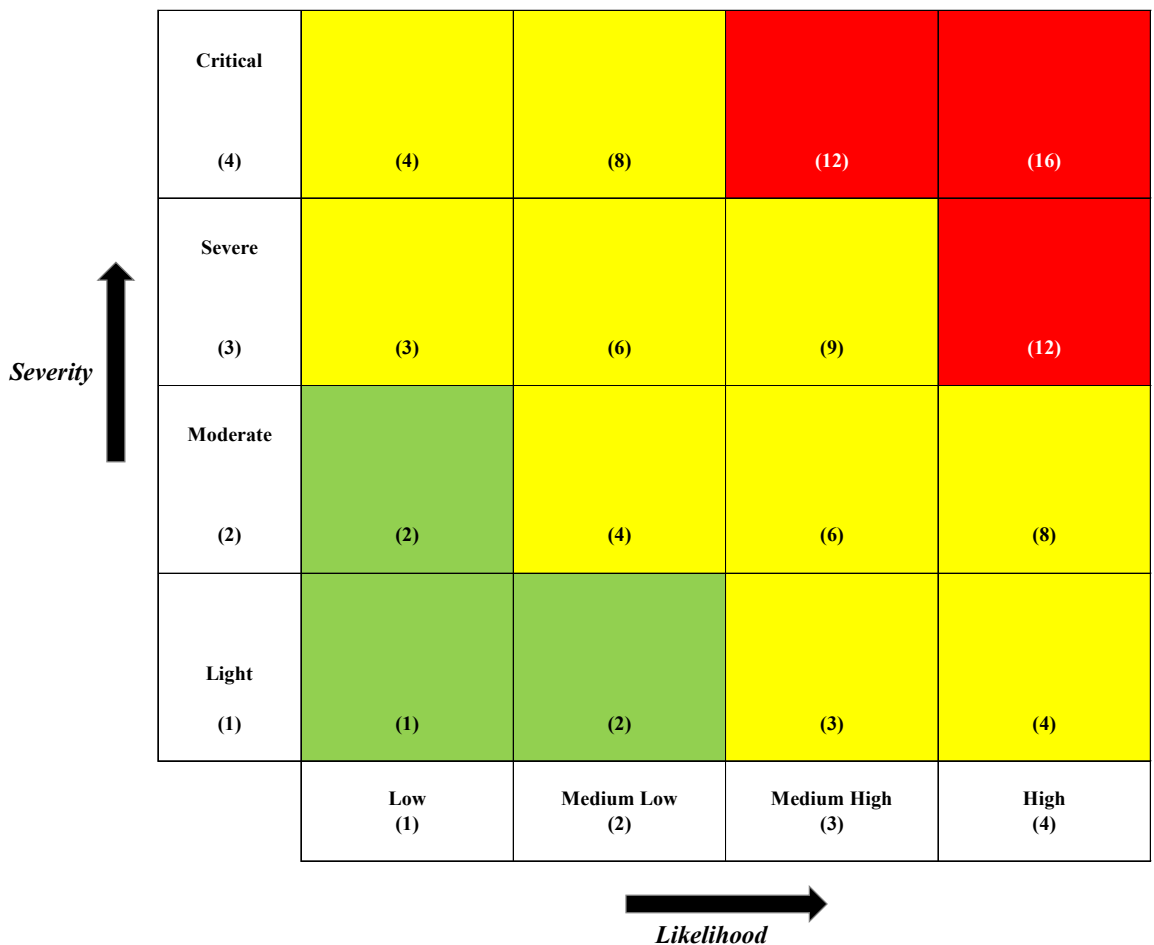

**Figure 1.** Cyber risk assessment matrix.

**Table 5.** Likelihood Definition.

| Score | Label | Likelihood Definition |
|---|---|---|
| **4** | **High** | - Attack can be performed remotely or with physical access to open ports and systems in the vessel.<br>- Can be performed with very minimal or no technical knowledge and with publicly available resources (e.g., Shodan, ExploitDB).<br>- Can be attacked from external network. |
| **3** | **Medium High** | - Attack can be performed with basic technical knowledge. |

| Score | Label | Likelihood Definition |
|---|---|---|
| | | - Can be performed with no change in exploits published online.<br>- Possible if the attacker is in either an internal or external network. |
| 2 | Medium Low | - Attack can be performed with moderate technical knowledge.<br>- Can be performed with minor changes in exploits published online.<br>- Possible if the attacker is in either an internal or external network. |
| 1 | Low | - Attack can be performed with advanced technical knowledge.<br>- Requires chaining of multiple exploits.<br>- Requires physical/remote access to the OT system where there is restricted access. |

**Table 6.** Severity Definition.

| Score | Label | Severity Definition |
|---|---|---|
| 4 | Critical | Consequences of a cyberattack on the vessel operations, crew, and the enterprise, such as:<br>- All operation systems, data, and resources unavailable, impacting safe operations.<br>- Impact on the lives of crew members e.g., collision, explosion, sinking of ship, imbalance of ship, loss of life.<br>- Economic loss to the enterprise. |
| 3 | Severe | Cyberattacks that lead to:<br>- Unauthorised access of the vessel network, system, data, and other resources that affect day-to-day vessel operations such as navigation, communication, propulsion, generators, cargo.<br>- Disruption of vessel network.<br>- Disruption of ship-to-shore communication link. |
| 2 | Moderate | Cyberattack prevents or impairs the normal authorised functionality of networks, systems, or applications by exhausting resources.<br>- Lack of availability of systems, data.<br>- Misleading communication between OT systems, causing damage to the ship, cargo and disrupting its operations. |
| 1 | Light | Suspected or potential unauthorised access to the vessel systems which leads to data breach (sensitive/non-sensitive information). |

Based on the risk scores determined with the 4x4 risk score matrix, as shown in Figure 1, the OT sub-systems and their associated cyber risks are classified into three categories based on their risk scores. The three categories are defined as follows: *High risk category (Risk Score: 12–16)*, *Medium risk category (Risk score: 3–9)* and *Low risk category (Risk score: 1–2)*. Such categorization facilitated us to define mitigation measures under each category in the form of a checklist.

*4.3. Checklist with Tiered Security*

Often, unsafe practices such as usage of default or weak passwords, outdated software or operating systems, etc., are some common causes that lead to cyber attacks. Hence, the most basic and crucial security controls must be implemented onboard the ships. Although there is no guarantee that a ship can be fully safeguarded from cyber threats, requisite actions can be taken as first line of defence against potential cyber threats.

In our guidelines we have provided a checklist, which is a list of mitigation actions that can be enforced in shipboard OT systems to manage the associated risks. To help maritime authorities and ship owners determine the cyber hygiene of their vessels, we introduce the concept of security tiers for listing the checklist. It consists of three tiers (Tier 1, Tier 2, Tier 3), where a tier of security refers to the *urgency of cyber risks* of a ship that ship owner needs to mitigate or manage.

The three security tiers are defined as follows:
- **Tier 1:** Includes cybersecurity measures for managing *high risk* cyber threats, which means that the measures stated under this tier are *highly recommended* to implement on-board vessels.
- **Tier 2:** Includes cybersecurity measures for managing *medium risk* cyber threats, which means that the security controls mentioned under this tier are *recommended* to have on-board.
- **Tier 3:** Includes cybersecurity measures for managing *low risk* cyber threats, which means that the measures listed under this tier are *good to implement on-board*, even though the risk level is low.

The checklist is designed to help ship owners and maritime authorities assess and improve the cyber hygiene of their vessels. The checklist can be easily enforced since the security measures are cost-effective and actionable.

## 5. Conclusions

Although adoption of emerging technologies in the maritime industry has become increasingly common, it opens up many attack surfaces for hackers to exploit. As OT systems play an important role in ship operations, it is crucial to protect them from the emerging cyber threats. In view of this, we developed guidelines that highlight the cyber risks and the impacts associated with OT systems, along with mitigation measures to manage those risks. The cyber risk assessment approach defined in the guidelines will aid in assessing the risks, whereas the checklist will guide in implementing appropriate security controls in shipboard OT systems. Thus, the guidelines will serve as a practicable guide for managing the cyber risks in shipboard OT systems and also ensure cyber hygiene in vessels.

Our guidelines have been disseminated by the Maritime and Port Authority of Singapore (MPA) to owners and operators of the Singapore Registry of Ships for their reference and use. An information paper "Voluntary cyber risk management guidelines for shipboard operational technology (OT) systems" has also been submitted by MPA to the 105th session of the IMO's Maritime Safety Committee, to make our guidelines freely available to all entities that may find them useful.

**Acknowledgements**
This work is carried out by iTrust, the Centre for Research in Cyber Security in Singapore University of Technology and Design (SUTD) and funded by the Singapore Maritime Institute (SMI) under the grant number SMI-2020-MA-04. The team would like to thank the Maritime and Port Authority of Singapore (MPA) for the support throughout this project. We would also like to extend our thanks to American Bureau of Shipping (ABS), Klynveld Peat Marwick Goerdeler (KPMG), Centre of Excellence in Maritime Safety (CEMS) in Singapore Polytechnic (SP), and the Singapore Shipping Association (SSA) for their valuable feedback on the draft version guidelines.